# PresenceSense: Zero-training Algorithm for Individual Presence Detection based on Power Monitoring

Ming Jin, Ruoxi Jia, Zhoayi Kang, Ioannis C. Konstantakopoulos, Costas Spanos
University of California, Berkeley, Berkeley, CA 94720
{jinming, ruoxijia, kangzy, ioanniskon, spanos}@berkeley.edu

## ABSTRACT

Non-intrusive presence detection of individuals in commercial buildings is much easier to implement than intrusive methods such as passive infrared, acoustic sensors, and camera. Individual power consumption, while providing useful feedback and motivation for energy saving, can be used as a valuable source for presence detection. We conduct pilot experiments in an office setting to collect individual presence data by ultrasonic sensors, acceleration sensors, and WiFi access points, in addition to the individual power monitoring data. PresenceSense (PS), a semi-supervised learning algorithm based on power measurement that trains itself with only unlabeled data, is proposed, analyzed and evaluated in the study. Without any labeling efforts, which are usually tedious and time consuming, PresenceSense outperforms popular models whose parameters are optimized over a large training set. The results are interpreted and potential applications of PresenceSense on other data sources are discussed. The significance of this study attaches to space security, occupancy behavior modeling, and energy saving of plug loads.

## 1. INTRODUCTION

Technological innovations in multiple disciplines extend the frontier of building science and opens up ample opportunities. First, low-cost manufacture of sensors and electrical meters reduces the economic barrier of the deployment of large sensor network. Second, pioneering works in the collection, communication, storage, and visualization of time series data, such as the Simple Measurement and Actuation Profile (sMAP) [3] and building-in-a-suitcase [4], greatly facilitate the profiling of energy consumption and building environments, as well as monitor-based commissioning (MBCx). Also recent development in statistical inference such as semi-supervised learning marks paradigm shift of using freely available unlabeled data for knowledge discovery [9].

Energy consumption of buildings, both residential and commercial, accounts for approximately 40% of all energy usage in the U.S. Plug loads alone represent 20% to 30% of the whole building energy use [1,2]. While top-down approaches like dynamic pricing is often effective for shared resources such as HVAC and lighting, considerable savings can be delivered by various bottom-up measures, including plug load metering, occupancy sensing, replacement of legacy equipment with Energy Star® equipment, and social games, where users have the central control of their devices [15]. Economic incentives and social motivations are effective means to induce behavior change.

Above all, reliable detection of individual presence in building space is a key component of an intelligent, occupant-friendly, and energy-efficient building. From the point of view of the building manager, it can help motivate occupants to save energy, scope occupants working behaviors to deliver personalized care and attention, as well as guarantee space security. For the occupants, the presence information can make them aware of their working patterns, identify occasional unusual behaviors, also get informed of energy consumption and ways to save energy. Cost effectiveness, nevertheless, is a major concern for the building owners given limited budgets. Additional presence sensors represent substantial amount of investment, which makes it not a practical solution for households or large commercial buildings.

It is, therefore, the objective of this paper to propose PresenceSense (PS), a framework that leverages existing infrastructure for presence inference and unusual behavior detection. In light of the submetering trends in green buildings, we base this framework on power measurement data in a typical office to infer individual presence as a replacement of additional presence sensor network that requires extra economic and set-up costs. The PresenceSense algorithm, as a semi-supervised learning method, does not require any training samples to avoid the labor-intensive labeling effort and it is very reliable. The paper is organized as follows. Section 2 introduces the experimental setup including several novel methods such as ultrasonic, acceleration, and WiFi, and the individual power monitoring system. The PresenceSense algorithm is discussed in Section 3. Section 4 reports the results of evaluating the algorithm in a typical office environment with diverse energy consumption profiles. In section 5, we provide an overview of previous works on occupancy detection and building plug-loads analysis. Conclusion and future works are provided in Section 6.

## 2. Experimental Procedure

This section provides a description of the presence sensor networks and the electric meter network in a typical office. Three novel types of sensors are implemented, namely ultrasonic sensor, acceleration sensor, and WiFi access points, which detects user presence by sensing the distance to sensor, movement of chairs, and presence of smart phones, other than the traditional motion detection. As illustrated in this section, each method has distinct false positive and false negative detection characteristics, and inevitably suffers from measurement noise. To evaluate these methods, we asked 4 users to provide a 5-minute resolution record of their presence in the office, with different markers for presence in the desk and presence in the office area except for his/her own desk. Their recording is used as a reliable ground truth in the evaluation.

## 2.1 Ultrasonic Sensor Network

Ultrasonic sensors measure the distance of the nearest obstacle, $d$, by recording the time it takes from sending to receiving the ultrasonic wave, $\Delta t$, according to the following relation:

$$d = \frac{1}{2} \cdot \Delta t \cdot v_{sound} \tag{1}$$

where $v_{sound} \doteq 340$ m/s is the velocity of ultrasonic wave travelling in the air.

The ZigBee networking protocol is adopted for communication, which features a network that is power efficient, ad-hoc, and self-organizing with no centralized control. The Tree network topology, as shown in Fig. 1, consists of ZigBee coordinator (ZC), ZigBee Router (ZR), and ZigBee End Device (ZED). Coordinator is the root of the tree and connects to the database through serial connection. It stores information about the network and acts as the Trust Center and repository for security keys. The ZigBee Router runs applications and acts as intermediate router. As parent of end devices, it acts as their mailbox, store messages while the end device is asleep and forward them when the end devices wakes back up. The protocol is automatically managed right inside the mesh network radios with no additional components or code required. ZigBee End Device is the sensor module that is placed on each person's desk. It only talks to parent node and cannot relay message.

The sensor module, or ZED, consists of an Arduino microcontroller, which controls the ZigBee module for communication and the ultrasonic sensors for reading. The ZigBee coordinator, namely base station, centrally coordinates all the sample collection by issuing requests periodically for each ZED one at a time. We use 10 seconds as the period to achieve a relatively high time resolution. Fig. 2 illustrates the ultrasonic measurement trace and the ground truth presence states.

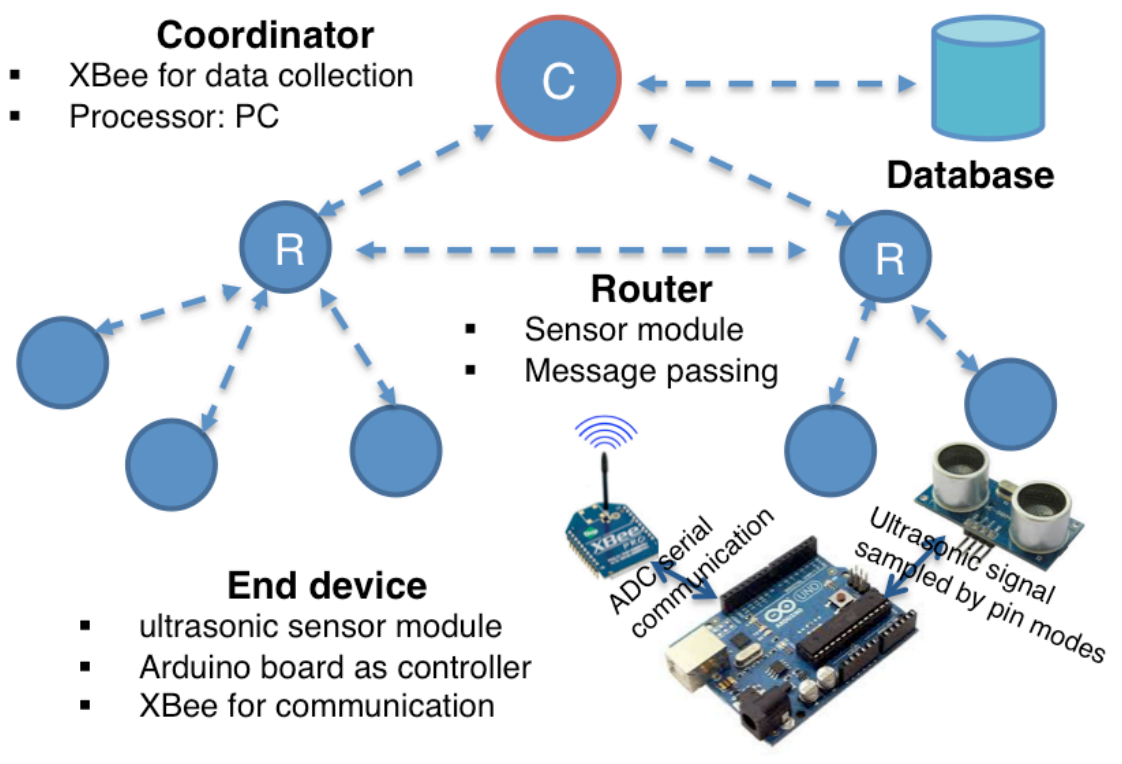

**Figure 1.** Network configuration of the ultrasonic sensor network based on the IEEE 802.15 standard. The sensor module is controlled by the Arduino microcontroller, and senses the distance by the ultrasonic sensor. The router is used for message passing and also as a sensor module. The coordinator collects the data periodically and stores in the local database.

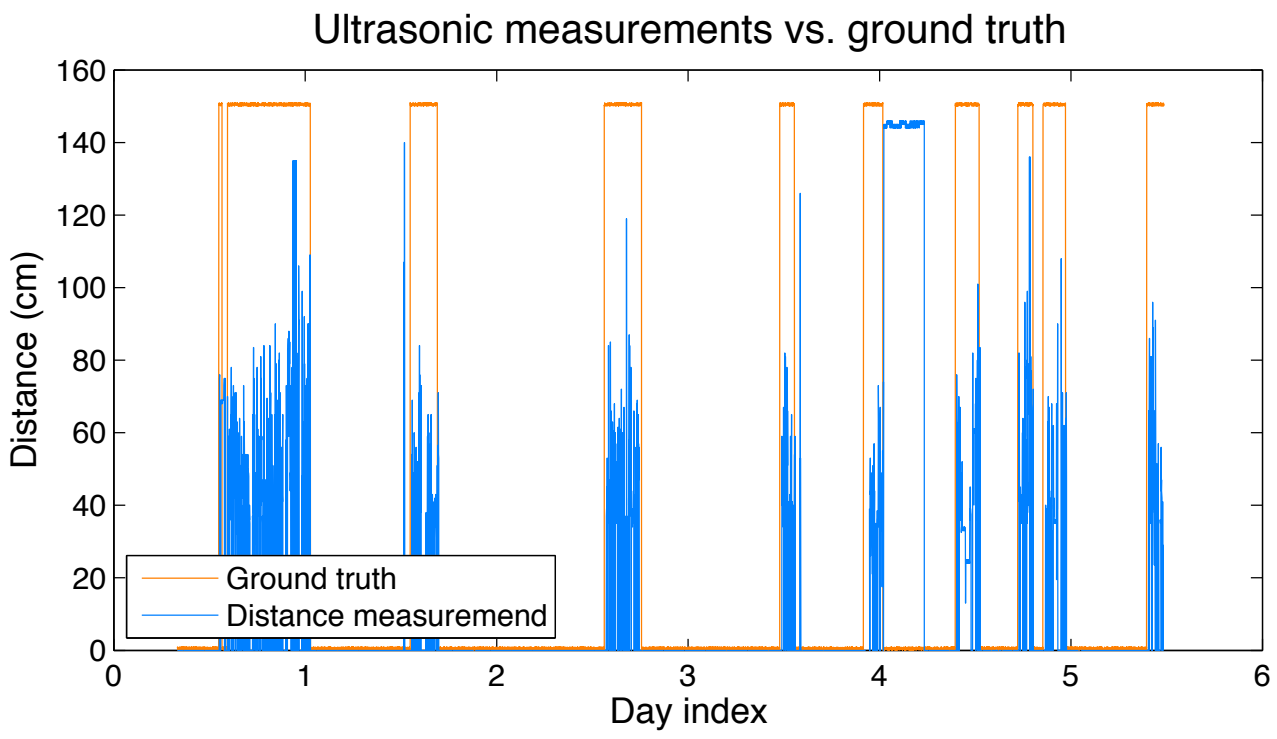

**Figure 2.** Typical ultrasonic measurements (blue) and the user presence ground truth (orange), where HIGH level indicates presence and LOW for absence. For some day the user might have left some obstacles in front of the sensor that the measured distance is in the range of normal presence but the change is not as large compared with presence.

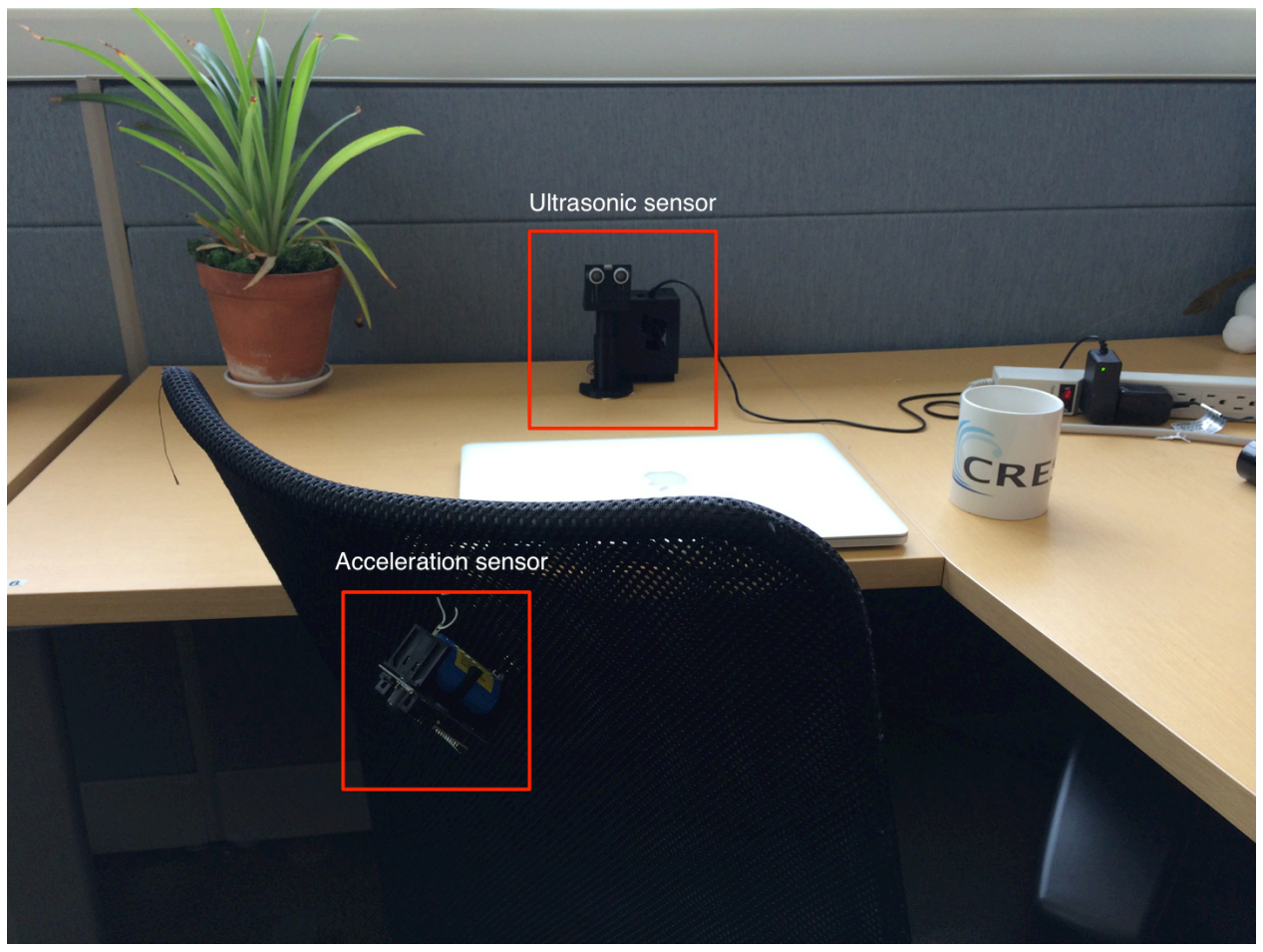

**Figure 3.** Experimental setup of the ultrasonic sensor and acceleration sensor. The location of the ultrasonic sensor is chosen so that it directly faces the user when he/she is working on the desk, and there should be no obstacles in front of it once the user departs. The acceleration sensor is attached to the chair in order to sense the chair movement. It is very uncommon that users share chairs, so the movement is a unique indicator of presence.

## 2.2 Acceleration Sensor Network

To provide ground truth for user presence, we made use of our previously developed low-cost, battery-powered Building-in-Briefcase (BiB) [4]. The device collects a rich set of environmental variables, such as temperature, humidity, ambient light, orientation sensing and motion detection. Furthermore, the small size of the sensor makes it possible to be installed easily in indoor environment. One BiB sensor is attached to the chair to detect motion by measuring the acceleration, as shown in Fig. 3.

All the BiB sensors attached to individual chairs are connected to centroid servers using WiFi. The acceleration measurements are sent to a local server and an Internet server simultaneously. All the data are stored on an on-board PostgreSQL server which also contains metadata about the incoming measurements. The internet server is similar to the local server. It uses an online cloud database to store all of the data.

The raw data includes the XYZ tri-axial acceleration measurements at a resolution of 1 second. To process the data, we obtain the standard deviation of the acceleration, $\sigma$ , given as

$$\sigma = \sqrt{\frac{1}{n}\sum_{i=1}^{n}\left(\bar{a}[i]-\mu\right)^2} \quad , \tag{2}$$

where

$$\mu = \frac{1}{n}\sum_{i=1}^{n}\bar{a}[i] \quad , \tag{3}$$

$$\bar{a}[i] = \sqrt{\left(a_X[i]\right)^2 + \left(a_Y[i]\right)^2 + \left(a_Z[i]\right)^2} \tag{4}$$

which is a popular feature based on acceleration data for activity recognition [19]. Fig. 4 demonstrates one typical trace of the standard deviation of acceleration data compared to the presence states. There are some days when the readings are noisy, which gives rise to deterioration of detection accuracy.

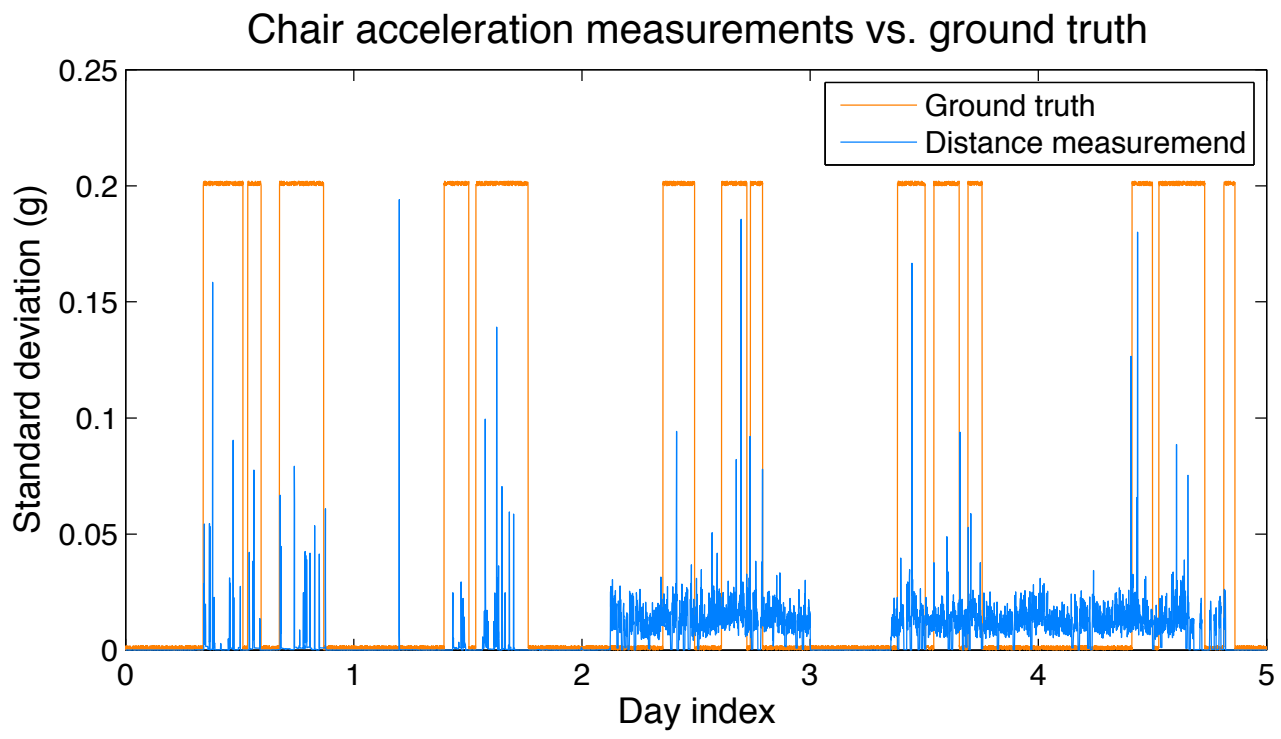

**Figure 4.** Chair acceleration (blue) and ground truth (orange) for user 17, where HIGH indicates presence and LOW indicates absence. For this user the threshold is chosen as .03g. For possible reasons of network or sensor degradation, the noise during the last several days is overwhelming, which deteriorates the performance of the presence detection accuracy.

## 2.3 WiFi Access Points

IEEE 802.11 (WiFi) is the most commonly used technology for internet access with widely available infrastructure in large numbers of commercial and residential buildings. For users with smart phones, it is likely that the WiFi transceiver is turned on. We employed a D-Link DIR-605L WiFi Cloud Router as the access points in the test-bed to detect the signal emitted from the phone. This method does not rely on the platform of the phone, and is relatively easy to implement. The collected WiFi Received Signal Strength (RSS) can be used as fingerprints to perform indoor positioning [20]. For the detection of presence, we only use the binary valued indicator, that if the connection of the phone to the access point is detected, it is indicated that user is present. Due to the IEEE 802.11 protocol, the communication is minimized to save energy, which means that the sampling period ranges from several seconds to tens of minutes. To obtain the presence inference, we apply a simple scheme that if at any time the access signal is detected, we denote points that are within 1 hour of the access point as presence.

## 2.4 Office Testbed

Our present study is carried out in the Center for Research in Energy Systems Transformation (CREST) located in Cory Hall on UC Berkeley campus, which is an office for graduate students in the EECS department. Each user works in a cubicle, where we install the ACme sensor to monitor the power consumption [17]. The power is measured at a resolution of one second. Connectivity to the Internet is provided via a small set of edge routers which function as a gateway to the Internet. The data is stored in our cloud database based on the Apache Cassandra.

The user is able to access his/her energy consumption by logging in our website (http://sbb01.eecs.berkeley.edu). It provides an interactive visualization of the real-time energy consumption to motivate users to monitor and save energy. The user is able to access his/her energy consumption by logging in our website, cloud database based on the Apache Cassandra.

By carefully observing the trends and distribution, we employ simple algorithms to infer the users' presence. For the ultrasonic data, as illustrated in Fig. 2, we design the interval of absence for each individual $i$ :

$$A_i = \left[a_1^i, a_2^i\right] \cup \cdots \cup \left[a_{k-1}^i, a_k^i\right] \tag{5}$$

and the decision rule is:

$$s^i(t) = \begin{cases} 1, \text{for } t \in \left\{n : d(n) \notin A_i\right\} \\ 0, \qquad \text{otherwise} \end{cases} \tag{6}$$

where $s^i(t)$ is the state of user $i$ , namely presence (1) or absence (0), at time $t$ . For the acceleration data, as shown in Fig. 4, we notice that high standard deviation of acceleration often indicates movement of user, and thus user presence. A simple threshold model as follow is applied

$$s^i(t) = \begin{cases} 1, \text{for } t \in \left\{n : \sigma(n) > \theta_i\right\} \\ 0, \qquad \text{otherwise} \end{cases} \tag{7}$$

where $\sigma(n)$ is given by Equ. (2). The WiFi data is smoothed to obtain the presence states as follow:

$$s^i(t) = \begin{cases} 1, \text{for } t \in \left\{n : \left|n - t_o\right| < \delta_i, t_o \in O_i\right\} \\ 0, \qquad \text{otherwise} \end{cases} \tag{8}$$

where $O_i$ is the observed time stamps when the access point detects the connection signal, $\delta_i$ is the time span for presence

As is shown in Table 1, these methods can only provide approximate inference results, as the measurement itself is inevitably noisy. To evaluate the accuracy, the traditional, labor-intensive method of actually recording each person's presence was also conducted. Given the difficulty and demanding requirements of this task, we only asked 4 users to participate in providing us with this information.

**Table 1. Accuracy of the various methods under investigation as evaluated against the ground truth. The accuracy is aggregated over all the users for ultrasonic, acceleration, and WiFi data, which are processed as illustrated above.**

| Methods | Absence | Presence | Total |
|---|---|---|---|
| Ultrasonic | 98.25% | 81.31% | 93.71% |
| Acceleration | 71.31% | 69.24% | 70.62% |
| WiFi | 90.13% | 47.43% | 77.21% |

## 3. ZERO-TRAINING ALGORITHM

In this section we introduce the main algorithm of PresenceSense, namely the zero-training algorithm, which require "zero" training labels for learning. Suppose the example space $\chi = \chi^1 \times \chi^2 \times \cdots \times \chi^v$ can be partitioned into $v$ views, namely feature spaces, and the class space is denoted as $\Upsilon = \{y_1, y_2\}$. The sets of labeled and unlabeled samples are given by $L = \left\{\left(\left(x_i^1,\ldots,x_i^v\right), y^i\right)_{i=1}^{|L|}\right\}$ and $U = \left\{\left(\left(x_i^1,\ldots,x_i^v\right), \cdot\right)_{i=1}^{|U|}\right\}$ respectively. Assume $v$ classifiers, $h_1,\ldots,h_v$ are trained based on each view of the example space.

The PresenceSense algorithm repeats the following steps until the maximum number of iterations is achieved or the stopping condition is met. At the start of the algorithm, all the samples are labeled according to the prior information incorporated in $h_1$, where $\chi^1$ is assumed to be a sufficient view as suggested by Blum and Mitchell [11]. Then all the other classifiers are trained using the initial labeled set, $L^1$. All the samples are relabeled by the majority voting rules, where ties are resolved according to the prior, which produce the new labeled set, $L^1_{new}$. The labeled set for the next iteration is updated according to the rule as detailed in Equ. (11), $L^{k+1} = g\left(L^k, L^k_{new}; \alpha_1, \alpha_2\right)$ as parameterized by the learning rates $\alpha_1, \alpha_2$. This constitutes one round of iteration. Detailed pseudo-code of the PresenceSense algorithm is provided in Table 2.

The finding of Angluin and Laird [13] on probably approximately correct (PAC) framework proposed by Valiant [10] is applied.

**Theorem 1 ([10]).** If we draw a sequence

$$m \ge \frac{2}{\varepsilon^2\left(1-2\eta_b\right)^2}\ln\left(\frac{2N}{\delta}\right) \tag{9}$$

samples from a distribution and find any hypothesis $L_i$ that minimizes disagreement with $\sigma$, where $\varepsilon$ denotes the hypothesis worst-case classification error rate, $\eta$ is the upper bound on the classification noise rate, $N$ is the number of hypotheses, and $\delta$ is the confidence, then the following PAC property is satisfied:

$$\Pr\left[d\left(L_i, L_*\right) \ge \varepsilon\right] \le \delta \tag{10}$$

where $d(,)$ is the sum over the probability of elements from the symmetric difference between the two hypothesis sets $L_i$ and the ground-truth $L_*$.

According to the proposed algorithm, the labeled set, $L_i^{k+1}$, where $i$ represents the label, is updated in each iteration according to the following rule:

$$L_i^{k+1} = \left\{L_i^k \cap L_{i,new}^k\right\} \cup Sample\left\{L_i^k \Delta L_{i,new}^k; \alpha_i\right\}, i \in \{1,2\} \tag{11}$$

where $L_i^k \Delta L_{i,new}^k$ is the symmetric difference, $\alpha_i$ is the sampling rate for the set of samples labeled $y_i$, for $i = 1,2$. In other words, the set of samples with label $y_i$ is given by the intersection of the previous set, $L_i^k$, and the labeled set produced by the classifiers of this round, $L_{i,new}^k$, together with a random subset of samples that these two sets do not share, with sampling rate of $\alpha_i$. If we denote $L_1^k = A_k \cup B_k$, where $A_k$ and $B_k$ are the set of correctly and incorrectly labeled samples respectively, and similarly $L_2^k = C_k \cup D_k$ where $C_k$ and $D_k$ are the set of correctly and incorrectly labeled samples respectively. Also denote $U_k = E_k \cup F_k$ where $E_k$ and $F_k$ are the set of unlabeled samples that should be labeled with $y_1$ and $y_2$ as the ground truth. For notation simplicity, we use $a_k, b_k, \ldots, f_k$ to denote the size of sets $A_k, B_k, \ldots, F_k$ respectively. Then the classification noise rate, $\eta_k$, is given by:

$$\eta_k = \frac{b_k + d_k}{a_k + b_k + c_k + d_k} \tag{12}$$

Assume the hypothesis makes a classification error independently for samples at the rate $\varepsilon^k$. We establish the following lemma to obtain an estimate of the classification noise rate $\eta_k$ and hypothesis classification error $\varepsilon^k$ in each iteration.

**Lemma 1.** The classification noise rate $\eta_k$ and hypothesis classification error $\varepsilon^k$ can be estimated assuming we have access to any two of the following quantities (categories do not matter):

a) *(prior information)* the number of $y_1$ in the samples, namely $a_k + d_k + e_k$, or $a_k + d_k$ for the labeled set

b) *(Type I or II error)* the misclassification rate of $y_1$ or $y_2$, namely $\frac{d_k}{a_k + d_k}$ or $\frac{b_k}{b_k + c_k}$

**Proof (Sketch).** According to the update rule in Equ. (11), the number of elements in the labeled set of the next iteration depends on the current iteration as follow:

$$a_{k+1} = a_k\left(1-\varepsilon^k\right) + \left(d_k + e_k\right)\left(1-\varepsilon^k\right)\alpha_1 \tag{13}$$

$$b_{k+1} = b_k\varepsilon^k + \left(c_k + f_k\right)\varepsilon^k\alpha_1 \tag{14}$$

$$c_{k+1} = c_k\left(1-\varepsilon^k\right) + \left(b_k + f_k\right)\left(1-\varepsilon^k\right)\alpha_2 \tag{15}$$

$$d_{k+1} = d_k\varepsilon^k + \left(a_k + e_k\right)\varepsilon^k\alpha_2 \tag{16}$$

Since we can observe the number of samples in $\left|L_1^k\right| = a_k + b_k$, $\left|L_2^k\right| = c_k + d_k$, and $\left|U^k\right| = e_k + f_k$, and also for those in round $k+1$, we can sum the pairs of Equ. (13,14), also (15,16). Together with two of the quantities proposed in Lemma 1, we can solve the system of equations for the estimation of $\varepsilon^k$ and $\eta_k$. □

**Remark (numerical solution):** The system of equations to be solved in Lemma 1 is non-linear, which makes it computationally costly to solve. Since the problem is defined for $0 \le \varepsilon^k \le 1$, we can perform a line search of $\varepsilon^k$. Given the value of $\varepsilon^k$, the system becomes linear and is very easy to solve by taking the inverse, or constrained quadratic programming. Then the optimal $\varepsilon^k$ that corresponds to the solution that best fits the remaining single equation should be chosen.

Theorem 1 provides the relationship among the number of training samples, $m$, and the classification noise bound, $\eta$, as well as the classification error rate, $\varepsilon$, of the hypothesis that minimizes the training error. Lemma 1 offers an estimation method of the classification noise rate $\eta_k$ in the $k$ round. Inspired by Zhou and Li [12] and Goldman and Zhou [14], we state the following corollary that guarantees the improvement of classification performance in each round of iterations.

**Corollary 1.** The gap between the learned and optimal hypotheses as shown in PAC property Equ. (10) is going to decrease with high probability in each iteration with suitable sampling rates, $\alpha_1$ and $\alpha_2$, whenever the following condition is satisfied:

$$\left(\left|L_1^{k+1}\right|+\left|L_2^{k+1}\right|\right)\left(1-2\eta_{k+1}\right)^2 > \left(\left|L_1^{k}\right|+\left|L_2^{k}\right|\right)\left(1-2\eta_{k}\right)^2 \quad (17)$$

where $\left(\left|L_1^{k+1}\right|+\left|L_2^{k+1}\right|\right)$ is the total number of samples in the training set in round $k+1$, and $\eta_{k+1}$ is the classification noise rate.

**Proof (Sketch).** Let $c = 2\mu \ln\left(\frac{2N}{\delta}\right)$ where $\mu$ is chosen to make the equality holds in Equ. (9), then we have $m_k = \frac{c}{\varepsilon_k^2\left(1-2\eta_k\right)}$, where $m_k = \left|L_1^{k+1}\right|+\left|L_2^{k+1}\right|$ is the number of samples in the labeled set. We introduce $u_k$ as follow for the simplicity of computation:

$$u_k = \frac{c}{\varepsilon_k^2} = m_k\left(1-2\eta_k\right)^2 \quad (18)$$

Since $u_k$ is proportional to $\frac{1}{\varepsilon_k^2}$, we have the following relation that if $u_{k+1} > u_k$, then $\varepsilon_{k+1} < \varepsilon_k$, and thus the corollary follows. □

**Remark (search of sampling rates).** As we have shown in Lemma 1, the number of elements in the labeled set is inherently a function of the sampling rates $\alpha_1$ and $\alpha_2$. By varying the sampling rates we can control the size of the labeled set in the next iteration, thus ensure the minimization of the optimal gap in each iteration. The calculation of sampling rates can be formulated as a feasible set problem in optimization.

## 4. Results and Discussion

The individual user's energy consumption monitoring started from July 31, 2013 till July 1, 2014. For the presence data, the approximate and ground truth are available from June 1, 2014 and June 18, 2014, respectively, till July 1, 2014. Table 3 is a summary of the profiles of subjects.

**Table 2. Pseudo-code of PresenceSense Labeling Algorithm**

```
PrsenceSense_Labeling(X, Prior, MaxIter)
  Inputs:  X: feature matrix of size n×v where n is the number
              of samples, v is the number of views
           Prior: expert knowledge used for initialization
           MaxIter: maximum number of iterations
  Initialization:
    L_1^0, L_2^0 ← Prior(X) #initial estimation by Prior
    stopCond ← false  #stop condition
    k ← 0 #iteration number
    η_0 ← 0.5  #classification noise rate
  Main program:
  while ¬stopCond ∧ k < MaxIter
    YMat ← emptyMatrix(n,v)
    for viewInd ∈ {1,...,v} do #train and predict in each view
      EstModl ← ModelEstimation(X^viewInd, L_1^k, L_2^k)
      YMat(•,viewInd) ← ModelPredict(EstModl, X^viewInd)
    end of for
    for sampInd ∈ {1,...,n} do #for each sample
      Y(sampInd) ← MajorityVote(YMat(sampInd,•))
    end of for
    L_{i,new}^k ← getSet(Y) #obtain the new labeled set
    L_i^{k+1} = {L_i^k ∩ L_{i,new}^k} ∪ Sample{L_i^k Δ L_{i,new}^k; α_i}, i ∈ {1,2} (*)
    η_{k+1} ← EstEta(L_{1,2}^k, L_{1,2}^{k+1}, α_1, α_2, estVals) (*) #Lemma 1
    stopCond ← checkStop(L_{1,2}^k, L_{1,2}^{k+1}, η_k, η_{k+1}) (*) #Corollary 1
    (Optional) search α_1, α_2 such that stopCond ← false by
    repeating (*) steps #line search with constraints
    k ← k+1 #update the iteration number
  end of while
  Outputs:  L_1, L_2 ← L_1^k, L_2^k # labeled sets for class 1 and 2
```

In the subsequent sections we first present result of feature explorations of power data, which are employed to generate multiple views in the example space, $\chi = \chi^1 \times \chi^2 \times \chi^3 \times \chi^4$. Then the PresenceSense algorithm is applied to multiple views to infer presence of each individual based on their power consumption. The classification results are interpreted and compared against ground truth.

**Table 3. User possession and usage of plug-loads. The user id is assigned to protect user identity in the experiments.**

| | Desktop | Monitor | Laptop | Lamp | Chargers |
|---|---|---|---|---|---|
| **6** | 0 | 0 | 1, often | 1 | 1-2 |
| **8** | 0 | 1 | 1, often | 1 | 1 |
| **17** | 1 | 1 | 1, seldom | 1 | 1-2 |
| **20** | 1 | 2 | 1, often | 1 | 1-3 |
| **26** | 1 | 1 | 1, often | 1 | 1-2 |

## 4.1 Power Feature Exploration

In this section features based on electricity consumption, including power level, edge effects, and rippling effects, are presented which are informative about user presence inference.

### 4.1.1 Power Level

Generally speaking the electricity consumption when the user is present is higher, since devices such as personal computer and/or laptops, desk lamp, electric chargers are turned on during working time. Fig. 5 illustrates power level distribution for the absence and presence states.

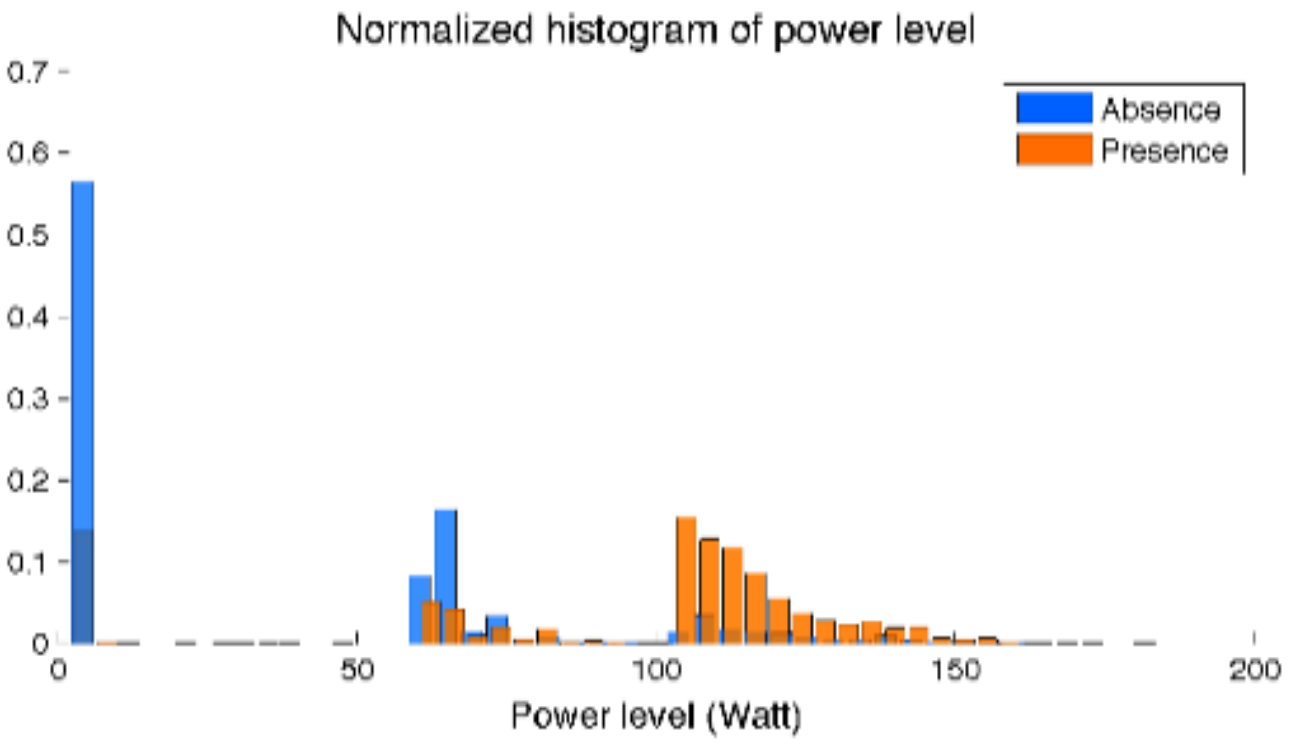

**Figure 5.** Normalized histogram of power level for user 17. The plug-loads profile includes a desktop, a monitor, a laptop, and various chargers. The non-zero power during absence is due to the desktop not being turned off when the user left. Simple absolute thresholding will not work in this case.

As can be seen, during the absence states, 2 distinct levels of demands exist, corresponding to desktop being left on or off when user is absent. The number of distinct levels ranges from one (1) to five (5) depending on the profile of devices in the cubicle. Also power level distribution during business hours, compared with non-business hours, exhibits wider spreads. Due to the behaviors of users who do not turn off their devices when absent, despite that power level is a valuable source of information, some other features are necessary to be acquired to improve the classification.

### 4.1.2 Edge Effect

The change of states usually happens with a large increase or decrease of power in a short amount of time, as marked by an edge in the trace of power. It is also observed that transition power does not belong to any stationary power distributions during presence and/or absence. Fig. 6 shows the power edge distribution during state changes.

As can be seen, edges are often associated with states change, though there are some exceptions when the user switches the states of the device during working. We design the maximum absolute change (MAC) to capture this edge effect:

$$MAC = \max_{1 \le i \le w} \left| x_i - x_{i-1} \right| \quad (19)$$

where $x_i$ is the power level indexed by $i$, and $w$ is the window size.

For highly autocorrelated signal, change of states is more robust to noise and also directly model the transition process, which are used in statistical models such as Hidden Markov Model (HMM). In addition, edge occurrence is related to hour of the day, which can be exploited by change-point detection methods [22].

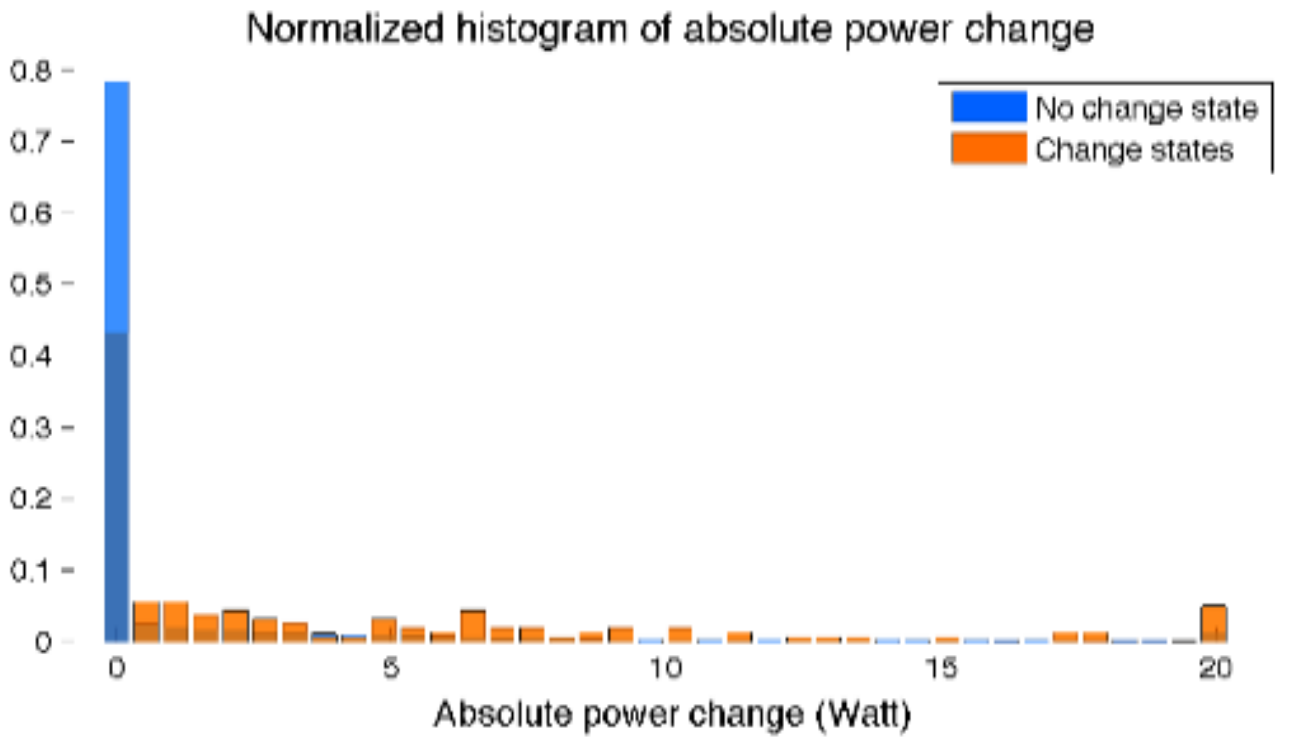

**Figure 6.** Normalized histogram of absolute power change. The change states refer to either departure or arrival. Since the value span is very large, all the values greater than 20W are counted at the boundary of 20W. The counts in each bin are divided by the total number of points in that category to obtain the normalized plot.

### 4.1.3 Rippling Effect

As one of the key observations, the rippling effect refers to the high frequency fluctuations exhibited in the power trace, as shown in Fig. 7 for an example.

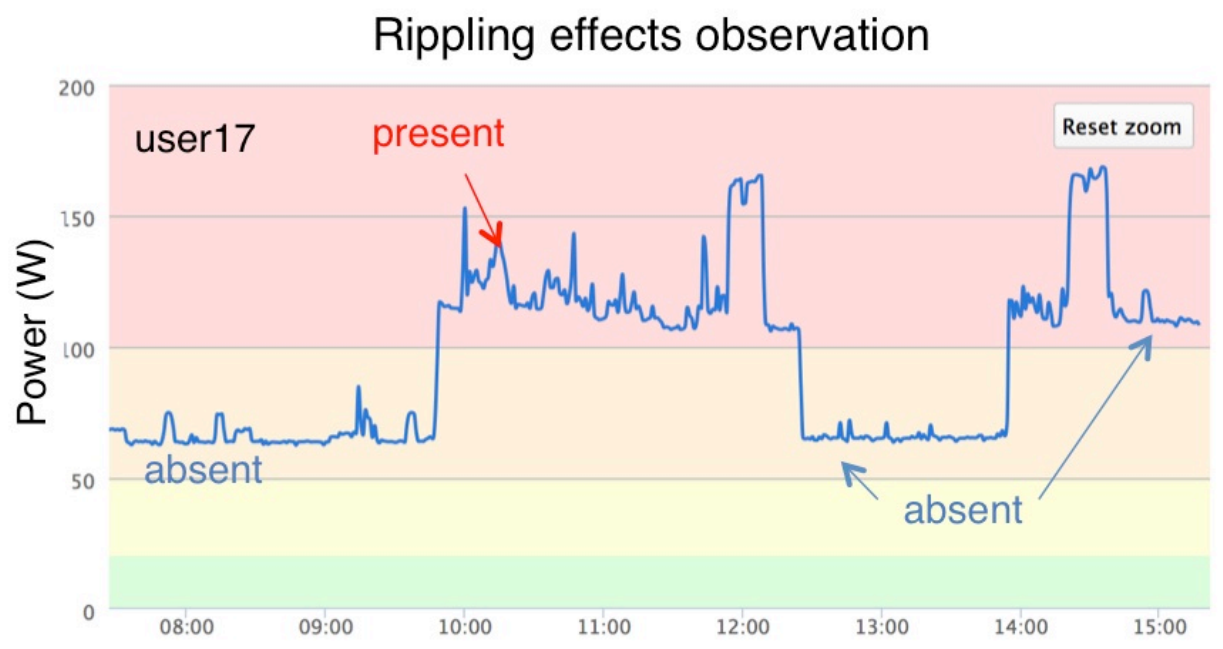

**Figure 7.** Observation of electricity consumption trace which reveals the correlation of rippling effects and user presence states. When the device is under usage, the intensity of power fluctuation increases, which is a useful indicator for presence.

Based on the rippling effect, some most informative and reliable metrics can be designed to capture the information from the high-resolution electricity measurement:

- Mean of absolute difference (MAD):

$$MAD = \frac{1}{w} \sum_{i=1}^{w} \left| x_i - x_{i-1} \right| \quad (20)$$

- Mean of absolute height difference (MAHD):

$$MAHD = \frac{1}{n-1} \sum_{i=1}^{n} \left| x_{c(i)} - x_{c(i-1)} \right| \quad (21)$$

- Standard deviation (SD):

$$SD = \sqrt{\frac{1}{w-1} \sum_{i=1}^{w} \left( x_i - \bar{x} \right)^2} \quad (22)$$

where $\bar{x}$ is the average power, , $n$ is the number of change points in the span, whose indices are given by $c(i) \in \{j : x(j) \text{ is local maximum or minimum}, 1 \le j \le w\}$. The conditional distribution of SD for presence and absence states is plotted in Fig. 8, and MAD and MAHD are very similar.

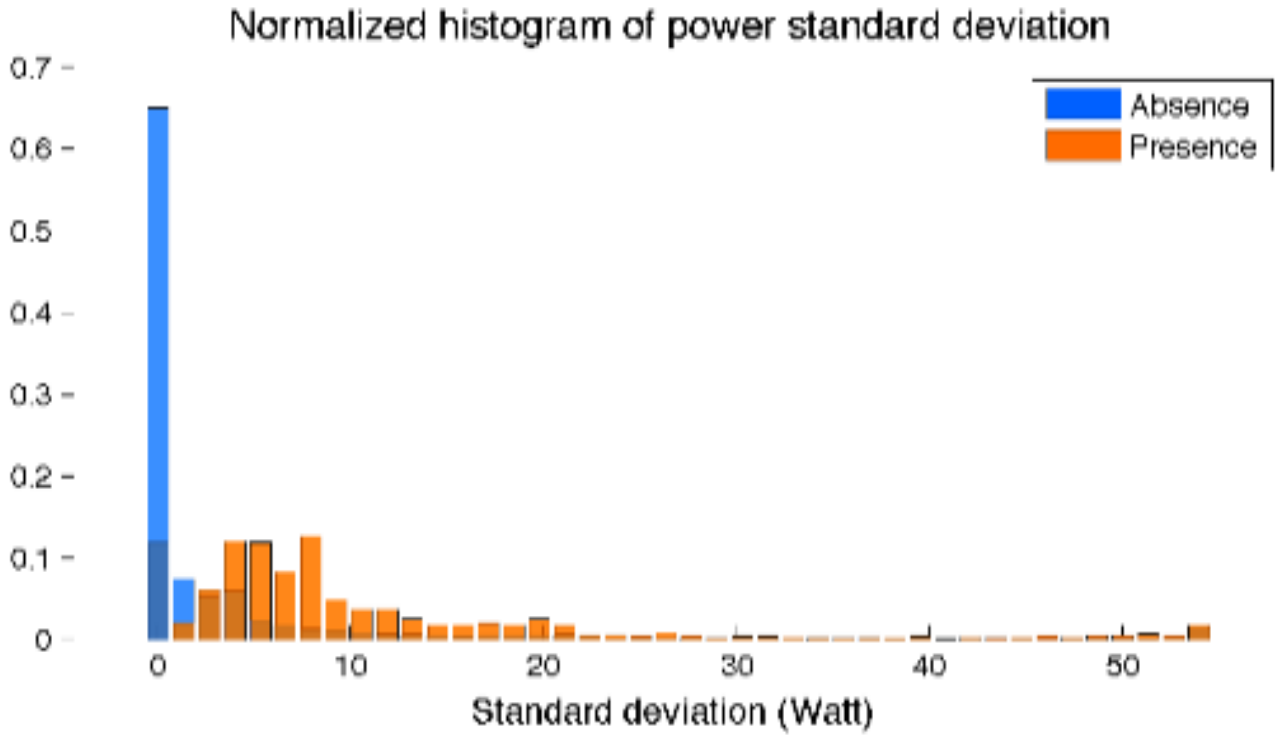

**Figure 8.** Normalized histogram of power standard deviation for user 17 as a measurement of the rippling effects. All the values greater than 55W are counted at the boundary for clarity.

As can be seen, all the features based on power rippling effects can achieve satisfactory separation of states, with SD being slightly better than the others. Based on experiences, to achieve a good balance of the aggregation of noise and user activity, 1min resolution is usually sufficient. It is worth to mention that rippling effect itself does not directly link to the states of being present or absent, as it indicates current intensity of device usage; therefore, it needs to be combined with other information to infer presence state.

## 4.2 PresenceSense Classification

In this section we present classification results as evaluated against the ground truth. As we propose in Section 3, PresenceSense is a zero-training algorithm, which means it does not require any tedious labeling of the data. Theorem 1 provides insights into the optimal learning process that we can allow a certain bounded classification noise in the training set as long as the number of training samples meets the requirement as set by Equ. (9). This result is used to guarantee the reliability of the learning result even with noisy training labels, where samples taken during absence are incorrectly marked as presence and vice versa.

The electricity consumption data is often correlated with each user's schedule, which can be used as useful prior knowledge for the PresenceSense initialization. Usually people are absent during non-business hours, which can be 0-8am for some users and 6pm to 8am for others. PresenceSense is not particularly demanding about the accuracy of the schedule, as long as the misclassification rate is less than 0.5. For most users they are absent during 8pm to 8am, 12 hours in total, which can already achieve a misclassification rate of less than 0.5. For our classification, *we apply the same working schedule for all the users*, namely, 0-9am absent, 9am – 8pm present, and 8pm – 0 absent. As we demonstrated next, the algorithm can get close to the ground truth even with this rough, and even incorrect, initial knowledge.

In each iteration we update the labels of the data by aggregating over the predictions of several classifiers with majority voting scheme, as shown in Table 2. The model can be of any choice that is suitable. Since we observe the distribution of features, such as power level, MAC, MAD and SD can be captured in a simple model with possibly few mixture components, we apply Bayesian classification rule by making the assumption that features are independent given the class, which is used for Naïve Bayes classifier for text categorization [23].

The optimal classification rule for each classifier is given by:

$$F(x_i^v) = \arg\max_c p(C=c)\, p(X^v = x_i^v \mid C=c) \tag{23}$$

where the conditional distribution, $\hat{p}(X = x \mid C = c)$ can be estimated by kernel density estimation

$$\hat{p}(X = x \mid C = c) = \frac{1}{n}\sum_{i=1}^{n} K_h(x - x_i) \tag{24}$$

where $n$ is the number of samples in class $c$, $K_h(x - x_i)$ is the kernel function. We use the Gaussian kernel, $K_h(u) = \frac{1}{h\sqrt{2\pi}} e^{-\frac{1}{2h^2}u^2}$, where $h$ is the standard deviation. Then the following majority scheme is applied to aggregate the views:

$$F_{maj}(x_i^{1,..,v}) = \text{median}\left(F(x_i^1),...,F(x_i^v)\right) \tag{25}$$

There are several reasons to apply early stopping rule in the training. First, early stopping when the convergence criterion is satisfied can save unnecessary computational power and improve the efficiency. Also, in the situation where few or even no training labels are available, stopping rules provide a guideline to track the performance of the algorithm and ensure that the algorithm achieves the optimal solution. Since by stochastically assigning training labels, it is possible that some misclassification will lead to confusion to the algorithm, which in turn tries to "correct" previously assigned accurate labels, the early stopping rules can prevent the training from dramatically increasing the misclassification rates.

We define the stopping metric $\phi_k$ as follow:

$$\phi^k = \left(|L_1^k| + |L_2^k|\right)(1 - 2\eta_k)^2 - \left(|L_1^{k-1}| + |L_2^{k-1}|\right)(1 - 2\eta_{k-1})^2 \tag{26}$$

where $|L_1^k|$ and $|L_2^k|$ are the number of training samples with label 1 and 2 at iteration $k$ respectively. The classification noise rate at iteration $k$, $\eta_k$, is estimated by Lemma 1. By Corollary 1, if the stopping metric is negative, then we don't have the guarantee that the misclassification rate will decrease in the next round. Therefore, it is recommended that the algorithm stop as soon as possible to avoid potential increase of misclassification rate.

We apply the PresenceSense algorithm on all the users' electricity consumption data. In general the early stopping indicator fires when the algorithm is in proximity to the optimal solution. As shown in Fig. 9 which shows the convergence of misclassification rate together with the stopping indicators, the indicator fires the first time after a sharp decrease of misclassification, then remains silent when the algorithm further improves incrementally until convergence. In this case, the optimal stopping time is around 3, but the iteration can also keep going until 30, but the improvement is minimal.

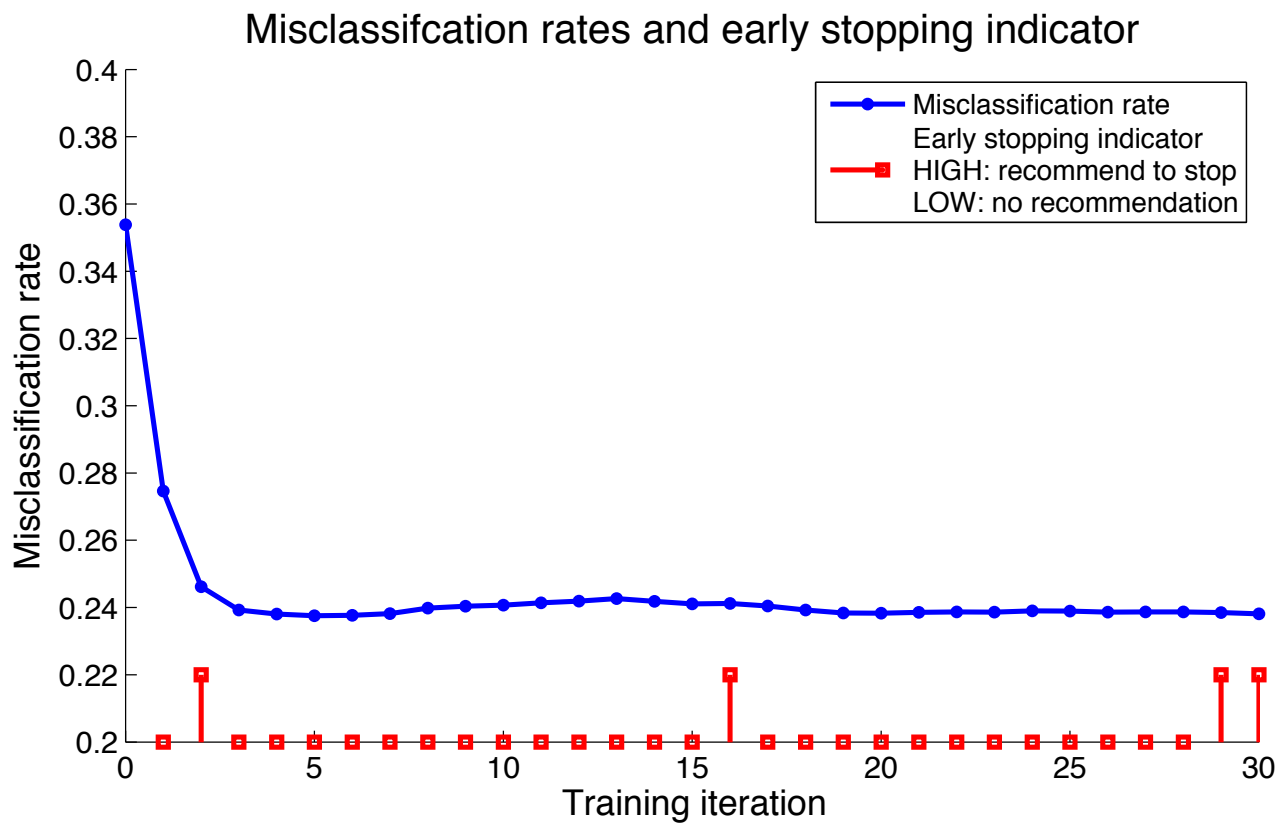

**Figure 9.** Misclassification rates and early stopping indicator for user 17. The misclassification rates drops significantly in the first several iterations, and converge as the iteration increases. The stopping indicator fires frequently when the algorithm is detected to be converging.

The importance of early stopping rule can be further appreciated in the case for user 8, whose plug-load profile does not include a desktop as shown in Table 3. As shown in Fig. 10, the PresenceSense achieves an optimal solution after only 2 iterations, and starts to include unnecessary and probably wrong labels afterwards. By inspection of the misclassification curve, the optimal stopping time is 2, which is also suggested by the stopping indicator. Since the misclassification rates deteriorate significantly afterwards, the indicator fires very frequently during the degradation, which strongly suggest that the process should be terminated. The profiles for the other two users are similar to the illustrated examples so they are not included here.

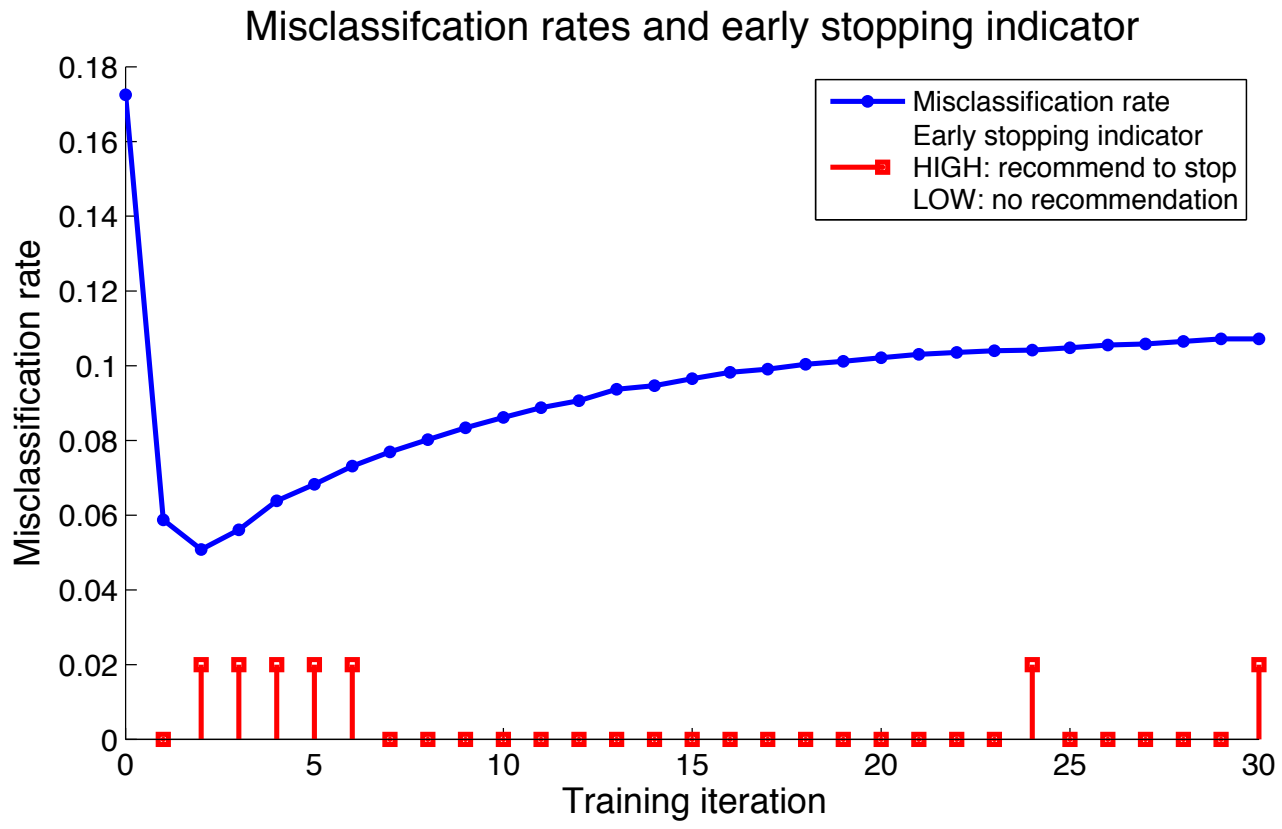

**Figure 10.** Misclassification rates and early stopping indicator for user 8. The misclassification rates drops significantly in the first several iterations, and starts to increase due to the random training errors. The stopping indicator fires frequently when this happens as a strong indication for termination.

As expected, the stopping indicator fires frequently when the misclassification rates increases significantly, and remains silent otherwise. It does not require any training labels or additional information, since the stopping metric shown in Equ. (26) can be computed readily with all the known information and Lemma 1. Therefore it is very convenient to work in practice and should be used whenever possible to ensure optimal solution. The accuracy of PresenceSense as compared with other popular models is listed in Table 4.

**Table 4. Accuracy of each model for different users.** The three numbers separated by '/' represent the detection rates given that the user is *absent*, *present*, and *overall*, respectively. The *best* performance is marked in bold, and the *second best* is underlined for comparison. Chg/Th: power level change threshold. Abs/Th: power level threshold. Prc/Th: change percentage threshold.

| | Chg/Th | Abs/Th | Prc/Th | PresSence |
|---|---|---|---|---|
| **8** | .87/.36/.72 | **.97**/.62/.86 | .79/.45/.69 | **.97/.71/.89** |
| **17** | .69/.64/.68 | **.92**/.76/.86 | .83/.49/.71 | **.92/.77/.87** |
| **20** | .67/.67/.67 | **.94/.69**/.86 | .87/.40/.72 | **.94**/.68/**.87** |
| **26** | .80/.14/.62 | **.99**/.66/.90 | .87/.15/.67 | .96/**.84/.93** |

The power level threshold model determines a threshold based on power level. If the power exceeds the threshold, the model infers presence. The change and percentage threshold models work in a similar way, except that the metric used are change and percentage in power respectively. When the change exceeds a threshold, if the state is absence, then it makes a transition to presence and vice versa. For the comparison, we use all the training labels for these models, and optimize over the space of threshold values to find the optimum. Even in this scenario, PS outperforms all the others in most cases.

The presence inference by PresenceSense and ground truth is shown in Fig. 11 for user 8 and 17. As we show in Fig. 12, the learned hourly schedule is very close to the ground truth, which means that PresenceSense can improve the estimation even with inaccurate initial knowledge.

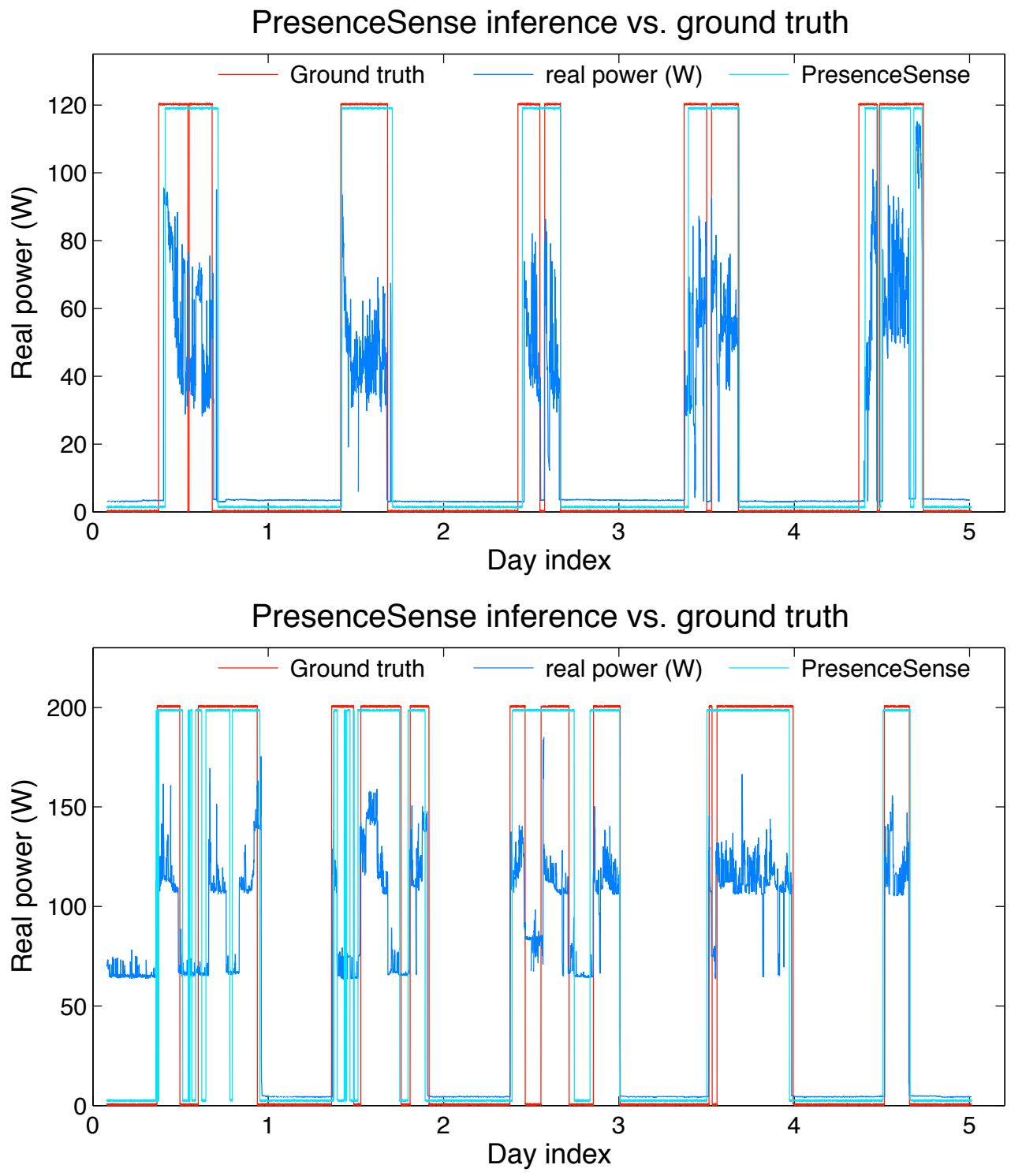

**Figure 11.** Presence inference by PresenceSense vs. ground truth for user 8 (top) and 17 (bottom). In addition to sensing the power level, PresenceSense captures the rippling effect which correlates with user presence to lower the misdetection rates for presence.

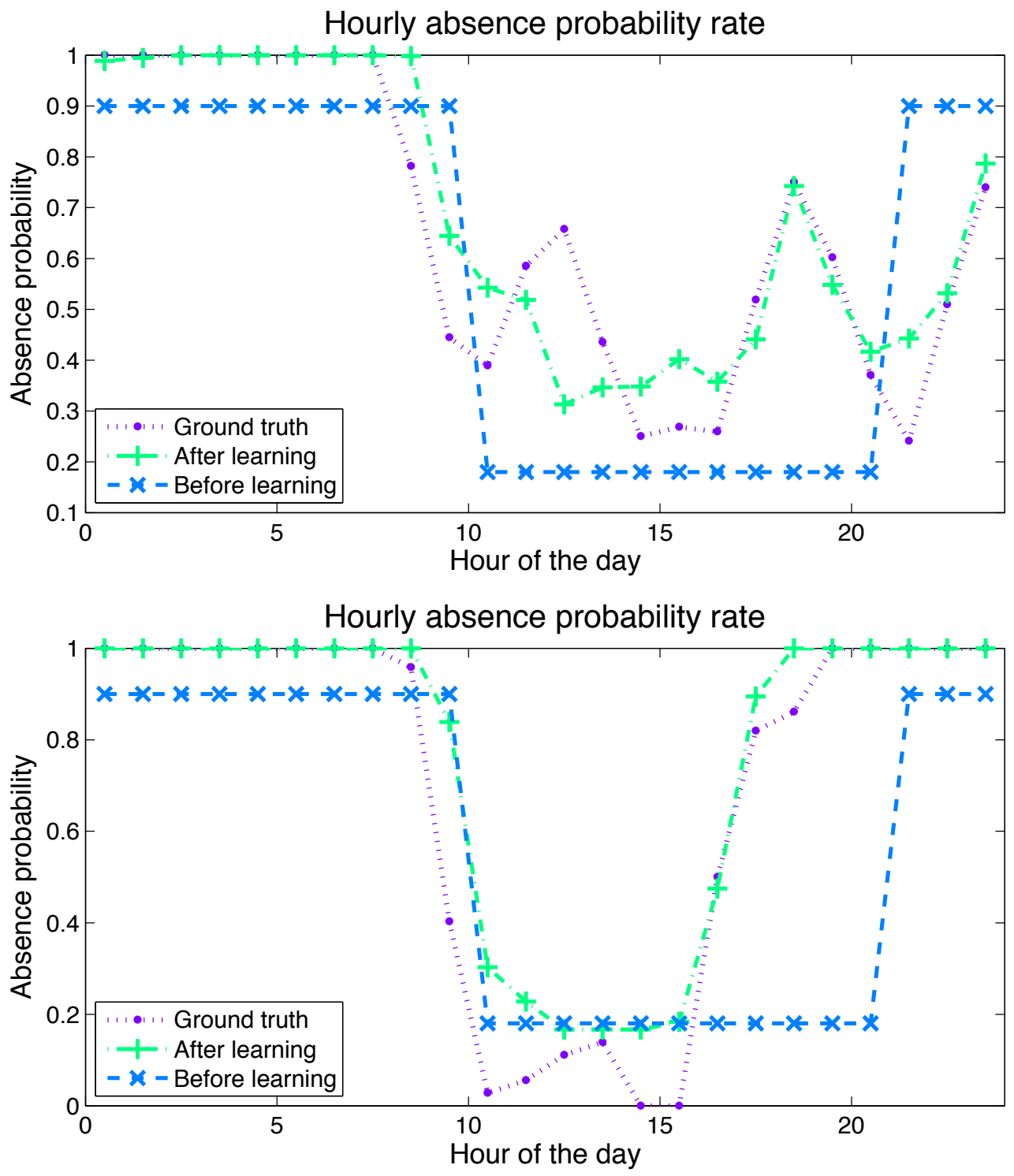

**Figure 12.** Comparison of hourly absence rates before learning, after learning, and ground truth for user 17 (top) and 8 (bottom).

# 5. RELATED WORK

This section provides an overview of previous works in presence sensors, occupancy sensing, and plug-in loads sensors. The difference of the current study is also highlighted.

*A. Presence Sensor and Occupancy Sensing*

Occupancy sensing is a natural extension of the presence sensing in indoor environment. Occupancy sensing endeavors to predict indoor occupancy level based on the correlation of sensor measurements to room occupancy.

Some of the sensors can infer the static occupancy information. For example, $CO_2$ sensor has been deployed to infer the number of people in a zone [5]. Although there is a response time of the sensor to human movements, careful calibration can be done based on PDE framework to achieve good modeling. Other sensors can infer the movement of occupants. Particulate Matter (PM) sensor is shown to correlate well with the human movement [6]. PIR sensors and video camera sensors have also been proved to have strong correlations with human activities [7,8], and have been used a set of experiment to infer occupancy in different zones in a building. When individual types of sensors fail to give reliable performance on occupancy, sensor fusion is used which integrate the results from different sensors. The authors in [26] proposed a sensor utility framework based on $CO_2$ sensors, plug-in load sensor, video camera sensor, and PIR sensors. They formulate a quadratic optimization framework based on mass conservation law.

*B. Plug-in loads Sensors*

Plug-in loads are believed to carry information on the building performance, sustainability and efficiency [1,16]. The ACme sensor has been developed by UC Berkeley to smartly measure the power consumption of plug-in loads, based on a resistance Power IC [17], and has been adopted in home energy sensing application [18]. LBNL used ACme system to measure a larger building space and demonstrated reliable performance [21]. The work from [26] showed a usage of power meters to help sense the occupancy of individuals. Chen et al. [24] and Kleiminger et al. [25] demonstrated the use of electricity consumption data obtained by smart meters to detect presence in residential buildings. However, none of those work tried to systematically explore the hidden relationship between plug-in loads and occupant behavior.

Our work is different from the previous works by considering the relationship of plug-loads electricity consumption and individual presence. The PresenceSense (PS) algorithm, based on semi-supervised learning theory, provides an easy method of learning the occupancy schedules without requiring any training labels, which is essential for portability and scalability to apply in other buildings with large crowds.

# 6. CONCLUSION

Presence detection is a key component in smart buildings to improve energy efficiency, occupancy comforts, and space management. Nevertheless, intrusive methods are often costly and difficult to implement due to privacy concerns. In this study we investigated the non-intrusive detection method based on individual power consumption. Since more and more buildings choose to closely monitor the plug-loads consumption as it consists of 20% to 30% of the total consumption, it does not require additional infrastructure investment. We also employed several presence detection methods, including the ultrasonic sensor, acceleration sensor, and WiFi access points, which represent some new possibilities of obtaining the presence states. Most importantly, we evaluated these methods against a set of ground truth that users supplied. The false positive and false negative rates for each method differ, as a reflection of the different characteristics.

We proposed PresenceSense (PS) algorithm to infer presence from the electricity consumption data. It is a zero-training algorithm, that is, it does not require any training labels which are usually costly to obtain. The algorithm works with very rough estimates of the user's working schedules, and iteratively relabels the data using the majority votes scheme by classifiers based on average power, power standard deviation, and absolute maximum power change. These features are demonstrated to have good separability for presence and absence states. The PresenceSense is compared with other common models, including the absolute power level threshold model, power level change threshold model, power level percentage change threshold model, whose threshold parameters are optimized over all the training sets. Even though the comparison is unfair for PresenceSense as it does not use any training sets, PS outperforms all the others in most of the cases as evaluated against the absence, presence, and overall detection rates. The theoretical effort to derive an early stopping rule is worthwhile as the early stopping indicator ensures that the algorithm finds the optimal sets of solution and saves computational power.

For future works, PresenceSense is useful for social games that are designed to motivate user energy saving [15]. Also unusual behaviors can be detected based on the presence inference, which are useful information for occupants and managers. The PS algorithm is based on semi-supervised learning theory, and it is of interests to improve the convergence bounds and estimates of classification accuracy, as it is essential when no training labels are available. Another interesting direction that we are exploring is to apply PS algorithm on other types of datasets such as indoor positioning and room-level occupancy estimation to achieve state-of-the-art results without training labels.

# 7. ACKNOWLEDGEMENTS

This research is funded by the Republic of Singapore's National Research Foundation through a grant to the Berkeley Education Alliance for Research in Singapore (BEARS) for the Singapore-Berkeley Building Efficiency and Sustainability in the Tropics (SinBerBEST) Program. BEARS has been established by the University of California, Berkeley as a center for intellectual excellence in research and education in Singapore.

# 8. REFERENCES

[1] Poll, Scott, and Christopher Teubert. "Pilot study of a plug load management system: Preparing for sustainability base." Green Technologies Conference, 2012 IEEE. IEEE, 2012.

[2] Moorefield, L., Frazer, B., Bendt, P. (2008). Office Plug Load Field Monitoring Report. Durango, CO: Ecos

[3] Dawson-Haggerty, S., Jiang, X., Tolle, G., Ortiz, J., & Culler, D. (2010, November). sMAP: a simple measurement and actuation profile for physical information. In Proceedings of the 8th ACM Conference on Embedded Networked Sensor Systems (pp. 197-210). ACM.

[4] K. Weekly, M. Jin, H. Zou, C. W. Hsu, A. M. Bayen, C. J. Spanos, Building-in-Briefcase, under review

[5] Weekly, Kevin, Nikolaos Bekiaris-Liberis, and Alexandre M. Bayen. "Modeling and Estimation of the Humans' Effect on the CO2 Dynamics Inside a Conference Room." arXiv preprint arXiv:1403.5085 (2014).

[6] Weekly, Kevin, et al. "Low-cost coarse airborne particulate matter sensing for indoor occupancy detection." Automation Science and Engineering (CASE), 2013 IEEE International Conference on. IEEE, 2013.

[7] Chenda Liao and Prabir Barooah. An integrated approach to occupancy modeling and estimation in commercial buildings. In American Control Conference (ACC), 2010, pages 3130{3135. IEEE, 2010.

[8] C. Liao, Y. Lin, and P. Barooah, "Agent-based and graphical modeling of building occupancy," Journal of Building Peformance Simulation, 2011.

[9] Semi-supervised learning. Vol. 2. Cambridge: MIT press, 2006.

[10] Valiant, L. G. (1984). A theory of the learnable. Communications of the ACM, 27(11), 1134-1142.

[11] Blum, A., & Mitchell, T. (1998, July). Combining labeled and unlabeled data with co-training. In Proceedings of the eleventh annual conference on Computational learning theory (pp. 92-100). ACM.

[12] Zhou, Z. H., & Li, M. (2005). Tri-training: Exploiting unlabeled data using three classifiers. Knowledge and Data Engineering, IEEE Transactions on, 17(11), 1529-1541.

[13] Angluin, D., & Laird, P. (1988). Learning from noisy examples. Machine Learning, 2(4), 343-370.

[14] Goldman, S., & Zhou, Y. (2000, June). Enhancing supervised learning with unlabeled data. In ICML (pp. 327-334).

[15] I. C. Konstantakopoulos, L. J. Ratliff, M. Jin, S. S. Sastry, , and C. Spanos, "Social game for building energy efficiency: Utility learning, simulation, and analysis," arXiv, Tech. Rep., 2014

[16] Gupta, Sidhant, Matthew S. Reynolds, and Shwetak N. Patel. "ElectriSense: single-point sensing using EMI for electrical event detection and classification in the home." Proceedings of the 12th ACM international conference on Ubiquitous computing. ACM, 2010.

[17] Jiang, Xiaofan, et al. "Creating greener homes with IP-based wireless AC energy monitors." Proceedings of the 6th ACM conference on Embedded network sensor systems. ACM, 2008.

[18] Jiang, Xiaofan, et al. "Design and implementation of a high-fidelity ac metering network." Information Processing in Sensor Networks, 2009. IPSN 2009. International Conference on. IEEE, 2009.

[19] M. Jin, H. Zou, K. Weekly, R. Jia, A. M. Bayen, Costas J. Spanos. "Environmental Sensing by Wearable Device for Indoor Activity and Location Estimation," arXiv:1406.5765 [cs.HC]

[20] Zou, H., Jiang, H., Lu, X., & Xie, L. (2014, March). An online sequential extreme learning machine approach to WiFi based indoor positioning. In Internet of Things (WF-IoT), 2014 IEEE World Forum on (pp. 111-116). IEEE.

[21] Dawson-Haggerty, Stephen, et al. "@ scale: Insights from a large, long-lived appliance energy WSN." Proceedings of the 11th international conference on Information Processing in Sensor Networks. ACM, 2012.

[22] Liu, S., Yamada, M., Collier, N., & Sugiyama, M. (2013). Change-point detection in time-series data by relative density-ratio estimation. Neural Networks, 43, 72-83.

[23] Schneider, K. M. (2003, April). A comparison of event models for Naive Bayes anti-spam e-mail filtering. In Proceedings of the tenth conference on European chapter of the Association for Computational Linguistics-Volume 1 (pp. 307-314). Association for Computational Linguistics.

[24] Chen, D., Barker, S., Subbaswamy, A., Irwin, D., & Shenoy, P. (2013, November). Non-Intrusive Occupancy Monitoring using Smart Meters. In Proceedings of the 5th ACM Workshop on Embedded Systems For Energy-Efficient Buildings (pp. 1-8). ACM.

[25] Kleiminger, W., Beckel, C., Staake, T., & Santini, S. (2013, November). Occupancy detection from electricity consumption data. In Proceedings of the 5th ACM Workshop on Embedded Systems For Energy-Efficient Buildings (pp. 1-8). ACM.

[26] Sean Meyn, Amit Surana, Yiqing Lin, Stella M Oggianu, Satish Narayanan, and Thomas A Frewen. A sensor-utility-network method for estimation of occupancy in buildings. In Decision and Control, 2009 held jointly with the 2009 28th Chinese Control Conference. CDC/CCC 2009. Proceedings of the 48th IEEE Conference on, pages 1494{1500. IEEE, 20